\documentclass[aps,prb,preprint,groupedaddress,showpacs,amsmath]{revtex4}
\usepackage{epic}
\usepackage{eepic}
\usepackage{amssymb,latexsym}
\usepackage{graphicx}

\newcommand{\SSpin}[2]{{\mathbf S}_{#1}{\mathbf S}_{#2}}
\newcommand{\Spin}[1]{{\mathbf S}_{#1}}
\newcommand{\al}{\alpha_{\rm L}}
\newcommand{\ar}{\alpha_{\rm R}}

\begin{document}

\title{Perturbation theories for the $S=1/2$ spin ladder with four-spin
ring exchange} 

\author{M. M\"{u}ller, T. Vekua, and H.-J. Mikeska}

\affiliation{Institut f\"{u}r Theoretische Physik,
Universit\"{a}t Hannover, Appelstrasse 2, 30167 Hannover, Germany}

\begin{abstract}
The isotropic $S=1/2$ antiferromagnetic spin ladder with additional
four-spin ring exchange is studied perturbatively in the strong
coupling regime with the help of cluster expansion technique, and by
means of bosonization in the weak coupling limit. It is found that a
sufficiently large strength of ring exchange leads to a second-order
phase transition, and the shape of the boundary in the vicinity of the
known exact transition point is obtained.  The critical exponent for
the gap is found to be $\eta\simeq1$, in agreement both with exact
results available for the dimer line and with the bosonization
analysis. The phase we determined after crossing the critical line is gapped and spontaneously dimerized. The results for the transition line 
from strong coupling and from weak coupling match with each other naturally. 
\end{abstract}


\maketitle

\section{Introduction}

At half filling and in the limit of small ratio
$x=t/U$ of hopping and on-site Coulomb repulsion the Hubbard model can
be mapped to an effective spin exchange Hamiltonian.  In the leading
order in $x$ the standard (bilinear) antiferromagnetic
nearest-neighbour Heisenberg exchange interaction with the exchange
constant $J=2t^2/U$ is obtained. Terms of higher order in $x$
yield, except bilinear exchange interactions beyond the nearest
neighbors, also exchange terms containing a product of four or more
spin
operators\cite{MacDonald88,MacDonald90,Takahashi77,EMueller01}.
Those higher-order terms were routinely neglected up to recent time,
when it was realized that they can be important for a correct
description of many physical systems.

For the first time biquadratic exchange was used for the description
of the magnetic properties of solid $^3$He \cite{Roger83}. Recently it
was suggested that some
strongly correlated electron systems like
cuprates\cite{Schmidt90,Honda93} and spin ladders \cite{MMueller99,Matsuda00}
are expected to exhibit ring exchange. The analysis
of the low-lying excitation spectrum of the p-d-model shows that the
Hamiltonian describing CuO$_2$ planes should contain a finite value of
ring exchange\cite{Schmidt90,Honda93,EMueller01,Coldea01,Katanin01}.

There is a number of experimental work like inelastic neutron scattering\cite{Eccleston98} and nuclear magnetic resonance\cite{Magishi98,Imai98} on Sr$_{14}$Cu$_{24}$O$_{41}$ and Ca$_8$La$_6$Cu$_{24}$O$_{41}$ as well as optical conductivity measurements\cite{Windt01,Nunner02} 
on (Ca,La)$_{14}$Cu$_{24}$O$_{41}$. All these substances contain spin ladders built of Cu atoms. 
The attempts to fit the experimental data without taking ring
exchange into account  yield an unnaturally large ratio\cite{Imai98} of $J_{\rm
leg}/J_{\rm rung}\approx 2 $ which is expected neither from the
geometrical structure of the ladder nor from electronic structure
calculations\cite{Mueller98}. It can be shown that 
inclusion of other types of
interactions in particular an additional diagonal interaction
does not remove this discrepancy \cite{Matsuda00}.

In the present paper we study the isotropic $S=1/2$ antiferromagnetic
spin ladder with additional four-spin ring exchange.  Starting from
the dimer limit of uncoupled rungs, we use the cluster expansion
technique to calculate the dispersion of the elementary excitations to
high order in the perturbation parameters $J_{\rm leg}$ and $J_{\rm
ring}$, and study the influence of the four-spin exchange on the
spectrum. In the $(J_{\rm leg}, J_{\rm ring})$
space we have found a transition boundary where the gap vanishes.  The
shape of this boundary is obtained in the vicinity of the exactly
known transition point lying on the ``dimer line'' where the exact
ground state is a product of dimers\cite{Kolezhuk98}. With the help of
Pad\'e-approximants we calculate the critical exponent $\eta$ for the
gap and obtain $\eta\simeq 1$, which is in agreement with the
exact results available for the dimer line.  We also analyze the
opposite limit of weakly coupled chains by means of the bosonization
technique and come to  the same conclusion on the linear behavior of the
gap. By this method we investigate the other side of the critical line. We show that the phase which emerges above the critical value of $J_{\rm ring}$ is a spontaneously dimerized phase with a finite gap to the elementary excitations.
This result is supported by recent numerical calculations.\cite{Troyer02,Hijii01}

\section{Model}

\begin{figure}[h]
\includegraphics[scale=1]{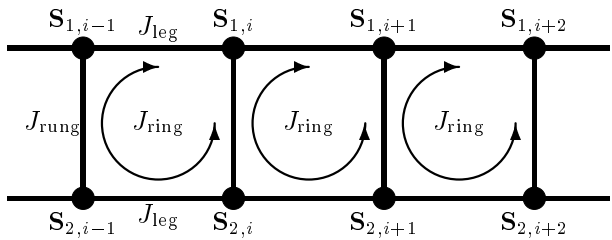}
\caption{Schematic structure of a two-leg ladder with additional ring
exchange}\label{fig:ring}
\end{figure}

We consider the isotropic $S=1/2$ antiferromagnetic spin ladder with
additional cyclic four spin exchange. Fig.~\ref{fig:ring} illustrates
the Hamiltonian, which is of the form
\[
\mathcal H = \mathcal H_{\rm rung} + \mathcal H_{\rm leg} + \mathcal H_{\rm ring}
\]
where
\begin{subequations}
\label{eq:ring}
\begin{eqnarray}
\mathcal H_{\rm rung} &=& J_{\rm rung}\sum_{i=1}^{{\mathcal N}}\SSpin{1,i}{2,i} \\
\mathcal H_{\rm leg}  &=& J_{\rm leg} \sum_{i=1}^{{\mathcal N}}\sum_{a=1,2}\SSpin{a,i}{a,i+1}\\ 
\mathcal H_{\rm ring} &=& \frac{J_{\rm ring}} 2 \sum_{\langle ijkl\rangle} \left(P_{ijkl} + P^{-1}_{ijkl} \right)
\end{eqnarray}
\end{subequations} 
In (\ref{eq:ring}) $\langle ijkl\rangle$ labels a four spin plaquette. $P_{ijkl}$ leads to a cyclic permutation of spin moments, i.e.
\begin{equation}
  P_{ijkl}\left|\begin{matrix} i & j \\ l & k\end{matrix}\right>
         =\left|\begin{matrix} l & i \\ k & j\end{matrix}\right>
  \quad\text{and}\quad
  P^{-1}_{ijkl}\left|\begin{matrix} i & j \\ l & k\end{matrix}\right>
              =\left|\begin{matrix} j & k \\ i & l\end{matrix}\right>.
\end{equation}
We rewrite the operator $P_{ijkl}$ as a product of two spin permutation operators and obtain the following result which contains both bilinear and biquadratic terms of the spin-$1/2$-operators:
\begin{eqnarray}
\label{Hring}
{\mathcal H}_{\rm ring} & =&
\frac{J_{\rm ring}}{2}
\sum_{\langle ijkl\rangle} \left[
\frac{1}{4} 
+ \Spin i \Spin j 
+ \Spin j \Spin k 
+ \Spin k \Spin l 
+ \Spin l \Spin i \right] \nonumber\\
&+&\frac{J_{\rm ring}}{2} \sum_{\langle ijkl\rangle} \Big[
\Spin i \Spin k 
+ \Spin j \Spin l 
\Big] \\
 &+& 2J_{\rm ring} 
\sum_{\langle ijkl\rangle} 
\Big[
(\Spin i \Spin j)(\Spin k \Spin l) 
 \nonumber\\
& +& (\Spin i \Spin l)(\Spin j \Spin k)  - (\Spin i \Spin k)(\Spin j \Spin l)
\Big]\nonumber
\end{eqnarray}
In further discussion the constant term is omitted, and 
periodic boundary conditions are used. 

On the so called ``dimer line'' where $J_{\rm ring}=J_{\rm leg}$ the ground state of the model (\ref{eq:ring}) is a product of singlet dimers placed on the ladder rungs.\cite{Kolezhuk98,MMueller99} Moreover, on this dimer line the propagating  triplet becomes an exact excitation,\cite{Kolezhuk98} and its energy is given by the following simple expression (the lattice constant is set to unity):
\begin{equation} 
\label{exact} 
\varepsilon_{t}(q)=J_{\rm rung}-2J_{\rm ring}+2J_{\rm ring}
\cos(q)\,. 
\end{equation}
It is easy to see that the gap of this excitation vanishes at $q=\pi$
for  $J_{\rm ring}=\frac 1 4 J_{\rm rung}$. It was
suggested,\cite{Kolezhuk98,MMueller99} that
 this point belongs to the line which has to be
identified  as corresponding to the
transition into the dimerized phase with spontaneously broken translational symmetry along the ladder. However, recent numerical calculations \cite{Honda01,Hikihara02} 
have created some doubt, indicating the possible
existence of a gapless phase on the other side of the transition line. 

\section{Elementary excitations and phase diagram}
\subsection{Rung dimer limit: Cluster expansion}
In this section we study the low-lying excitations of the above
model by perturbation theory. Therefore we start in the dimer limit where
$J_{\rm rung}$ is the only non-vanishing exchange constant and measure all
other interactions in units of $J_{\rm rung}$. In fact we have only two 
coupling parameters: 
\begin{equation}
\alpha_{\rm L}=J_{\rm leg}/J_{\rm rung}\quad\text{and}\quad
\alpha_{\rm R}=J_{\rm ring} /J_{\rm rung}.
\end{equation}

The dispersion is a function of both $\ar$ and $\al$ possessing
following form:
\begin{equation}
\omega(q)/J_{\rm rung} = \sum_{n=0}^{\infty}\sum_{i=0}^n 
\sum_{j=0}^n c^{(n)}_{i,j} \alpha_{\rm L}^j\alpha_{\rm R}^{n-j}\cos(iq).
\end{equation}

The dispersion was obtained by using the  cluster expansion formulation
\cite{singh90,gelfand95}
of the perturbation theory in $\ar$, $\al$. Clusters 
with maximum of $14$ edges were considered, which is an improvement
of $11$ orders to our previous work \cite{MMueller99}. Up to the third order
one has
\begin{eqnarray}
\label{eq:ringdisp}
\omega^{(3)}(q) &=& \mu_0 + \mu_1\cos(q) + \mu_2\cos(2q) +
\mu_3\cos(3q)
\nonumber\\
\mu_0 &=& 1 - 2\ar + \frac 3 8  \left( \al-\ar \right)^{2}\left( 2+ \ar+5\,\al \right) \nonumber\\
\mu_1 &=& \ar + \al - \frac 1 4  \left( \al-\ar \right)^{2}\left( \ar+\al \right)\\
\mu_2 &=&  - \frac 1 4  \left( \al-\ar \right)^{2}\left(1+ \ar+\al
\right) 
\nonumber\\
\mu_3 &=&  \frac 1 8  \left( \al-\ar \right)^{2}\left( \ar+\al \right).
\nonumber
\end{eqnarray}
One may notice that for $\al=\ar$ only the first-order term is left in
the expansion, and one recovers the exact result (\ref{exact}).

\begin{figure*}
\includegraphics[scale=0.45]{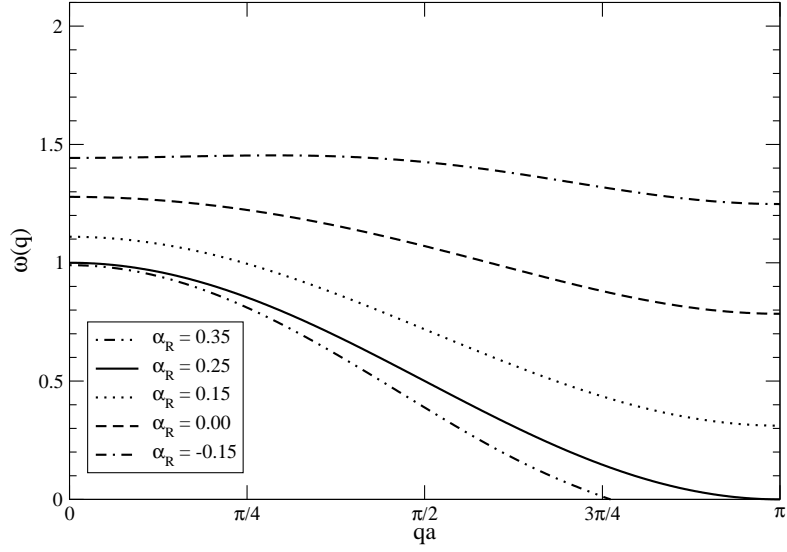}
\caption{\label{fig:dispersionL25} Dispersion for $\al=0.25$ up to the 14$^{\rm th}$ order and for various $\al$. On the exact line ($\ar=0.25$) the gap vanishes. Further increase of $\ar > 0.25$ leads to a loss of convergence.}
\end{figure*}

\begin{figure*}
\includegraphics[scale=0.45]{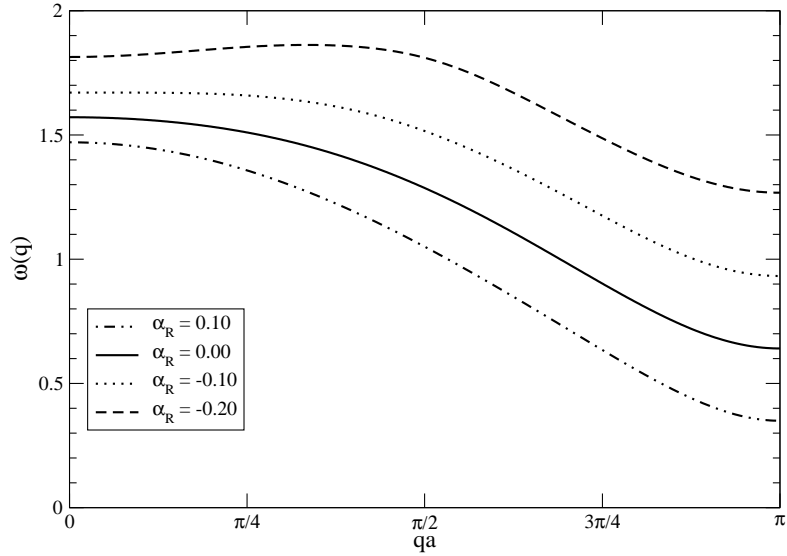}
\caption{\label{fig:dispersionL50} Dispersion for $\al=0.50$ up to the 14$^{\rm th}$ order and for various $\al$.}
\end{figure*}

In Fig.~\ref{fig:dispersionL25} and Fig.~\ref{fig:dispersionL50} we
plot some typical dispersions for fixed $\al=0.25$ ($\al=0.50$)
and varying $\ar$.  One can see that the gap at $q=\pi$ decreases with
increasing $\ar$ and finally vanishes. It is also seen that the
convergence of the method becomes worse when $\al$ and $\ar$
increase. However, the results from direct series expansion can be
improved by using the Pad\'e-approximation technique.

We have studied the vicinity of the exactly known transition point
$\al=\ar=0.25$ and
calculated the phase boundary where the gap closes. The resulting view
of the phase diagram is shown in Fig.~\ref{fig:phase}(a). We were not
able to proceed beyond the intervals which are marked by the arrows by
increasing the order. It seems that we hit the convergence radius of the
present method. 

We have also calculated the critical exponent $\eta$ of the
gap $\Delta_{\rm ring}$ as a function of both $\al$ and $\ar$ 
where we use the following definition:
\begin{eqnarray}
\Delta_{\rm ring} &\propto& |\al-\al^c|^{\eta(\ar)}\quad\text{and}\\
\Delta_{\rm ring} &\propto& |\ar-\ar^c|^{\eta(\al)}.
\end{eqnarray}
Within the convergence 
interval $\al=[0.17,0.36]$ resp. $\ar=[0.24,0.38]$ this exponent
is equal to $1$ with the accuracy of $\pm 0.01$. This is in agreement with
the exact result (\ref{exact}) and shows that the picture of the phase
transition does not change when one moves away from  the exact point
$\al=\ar=0.25$. This behavior of the gap indicates that the phase on the 
other side of the phase boundary is gapped. Indeed, linear growth of the 
gap indicates presence of a
relevant operator, whose amplitude changes sign at the
boundary. Existence of the high-$\ar$ gapless phase would, in
contrast, imply presence of a marginal operator which becomes
irrelevant on the boundary. That would necessarily mean
exponentially slow growth of the gap, which contradicts to our
observations. Those numerical results are also in a good agreement
with the conclusions of bosonization analysis (see the next section),
which predict a linear behavior of the gap at both sides of the transition.


\subsection{Decoupled chains limit: Bosonization}

In this section we perform the weak-coupling analysis ($J_{\rm leg}
\gg J_{\rm ring},\, J_{\rm rung}$) of our model. In a weak coupling-bosonization analysis we find a critical line separating rung singlet phase from spontaneously dimerized phase. We conclude that the critical line as well as the phases it separates match with each other in weak and strong coupling regimes. We show that the effect of the ring exchange in the infrared limit is reduced (up to marginal corrections)
to that of the leg-leg
four-spin interaction as considered earlier by Nersesyan and
Tsvelik\cite{Nersesyan97} in the context of spin-phonon coupling.
\par We write the spin operators on each chain in terms of their smooth and staggered parts:
\begin{equation} 
{\mathbf S}_{1,2}(x)={\mathbf J}_{1,2}(x)+(-1)^x {\mathbf n}_{1,2}(x).
\end{equation}
The effect of rung interaction is well understood by bosonization using Majorana fermion formalism \cite{Shelton96}. Now we will bosonize ring exchange, treating it as a perturbation to decoupled chains and at the end we will add bosonized-refermionized terms coming from rung interaction as in \cite{Shelton96}. We decompose the Hamiltonian in the following way:
\begin{equation}
H(x)=H_1(x)+H_2(x)+H_{\rm quad}(x)+H_{\rm biquad}(x)\,\, ,
\end{equation}
where $H_{1,2}$ are critical Gaussian models describing first and
second decoupled chains. $ H_{\rm quad}$ stands for the quadratic spin
interactions and $H_{\rm biquad}$ for the four-spin interactions
originating from the ring exchange term.  

Neglecting renormalization of the
intrachain interaction we first analyze the quadratic spin interactions
which can be cast in the following form:
\begin{eqnarray}
 H_{\rm quad} &\sim& (J_{\rm ring}^{\bot}+J_{\rm ring}^{\times}){\mathbf J}_1(x)
{\mathbf J}_2(x)
\nonumber\\
&+&(J_{\rm ring}^{\bot}-J_{\rm ring}^{\times}) {\mathbf n}_1(x){\mathbf n}_2(x) 
\end{eqnarray}
where $J_{\rm ring}^{\bot}=J_{\rm ring}^{\times}=J_{\rm ring}$.  Since
the scaling dimension of the smooth part of the spin operator is 1, while the dimension of the staggered part is $1/2$, no relevant terms are generated from the quadratic part of the ring exchange, and only marginal terms are left.
After bosonizing the biquadratic part only leg-leg interaction will
survive, because diagonal-diagonal and rung-rung terms give
non-distinguishable relevant contributions in the infrared limit
which cancel each other due to the overall opposite  signs in front of
them (which is fixed by the structure of the ring exchange):
\begin{equation}
H_{\rm biquad} \sim (J^{\rm RR}_{\rm ring}-J^{\rm DD}_{\rm
ring}+J^{\rm LL}_{\rm ring}) \epsilon_1(x) \epsilon_2(x), 
\end{equation}
where $\epsilon_{1,2}(x)=(-1)^x{\mathbf S}_{1,2}(x){\mathbf
S}_{1,2}(x+a_0)$ represents the dimerization operator of the first and
second chain, respectively,  and
$J^{\rm RR}_{\rm ring}=J^{\rm DD}_{\rm ring}=J^{\rm LL}_{\rm ring} =
2J_{\rm ring}$.  
In the Majorana representation (retaining only relevant
operators) we arrive at the following Hamiltonian:
\begin{equation}
H=\sum_{a=0}^3\int dx \left[ \frac{-iv}{2}(\xi_R ^a\partial_x \xi_R ^a 
-\xi_L ^a\partial_x \xi_L ^a)-im\xi_R^a \xi_L^a \right] 
\end{equation}
with $m=-c^2\alpha J_{\rm ring}/ 2 \pi$. $c$ is the
Lukyanov-Zamo\-lod\-chi\-kov constant at the $SU(2)$ AFM point and $\alpha$
is a non-universal, cutoff-dependent positive constant. Thus in the
weak-coupling limit we have effectively reduced the ring exchange to
the leg-leg interaction. The only difference between the bosonized
forms of the ring exchange and the pure leg-leg coupling stems from the
marginal current-current interaction which does not appear in the
leg-leg biquadratic interaction. In contrast to the
leg-leg interaction ring exchange is not invariant under independent global $SU(2)$
rotations of spins on each chain, and thus should not enjoy full
$O(4)$ symmetry. This symmetry is in fact lowered by marginal
operators. The refermionized version of the marginal current-current
interaction contained in ring exchange will take the following form 
 in the Majorana representation:\cite{Shelton96}
\begin{eqnarray}
H_{\rm marg}&=&J_{\rm ring}a_0\int dx
[ (\xi_R^1 \xi_L^1)(\xi_R^2 \xi_L^2)\nonumber\\
&+&(\xi_R^2 \xi_L^2)(\xi_R^3 \xi_L^3)+(\xi_R^1 \xi_L^1)(\xi_R^3 \xi_L^3)\nonumber\\
&-&(\xi_R^1 \xi_L^1+\xi_R^2 \xi_L^2+\xi_R^3 \xi_L^3)(\xi_R^0 \xi_L^0)]
\end{eqnarray}
Renormalizing the masses it weakly splits the $O(4)$ quadruplet into
a triplet and a singlet, consistent with the symmetries of ring exchange:
\begin{eqnarray}
\frac{m_t} m &\longrightarrow& 1+\frac{J_{\rm ring}a_0}{\pi v}\ln\frac{J_{\rm leg}}{|m|},\nonumber\\ 
\frac{m_s} m &\longrightarrow& 1-\frac{3J_{\rm ring}a_0}{\pi v}\ln\frac{J_{\rm leg}}{|m|}.
\end{eqnarray}
The full refermionized model, including both rung and ring exchange,
in the Majorana representation reads as: 
\begin{eqnarray}\label{eq:mass}
H&=&\sum_{a=1,2,3}\int dx \left[ \frac{-iv_t}{2}(\xi_R ^a\partial_x \xi_R ^a 
-\xi_L ^a\partial_x \xi_L ^a)-im_t\xi_R^a \xi_L^a \right] \nonumber\\
&-&\int dx \left[ \frac{iv_s}{2}(\xi_R ^0\partial_x \xi_R ^0 
-\xi_L ^0\partial_x \xi_L ^0)-im_s\xi_R^0 \xi_L^0 \right]
\end{eqnarray}
where
\begin{eqnarray}
m_t &=&\frac{c^2}{2\pi} (J_{\rm rung}-\alpha J_{\rm ring}),\nonumber\\
 m_s&=&-\frac{c^2}{2 \pi}(3J_{\rm rung}+\alpha J_{\rm ring})
\end{eqnarray}
From the above formulae, one readily obtains the line where the triplet mass vanishes. 
According to Ref.\ \onlinecite{Nersesyan97}, on
this fine-tuned line a phase transition from conventional Haldane
phase (rung exchange dominated phase) where spectrum displays coherent
single-particle (magnon) excitations to non-Haldane spontaneously
dimerized phase (ring exchange dominated phase) without coherent
magnon modes takes place. This transition belongs to the universality
class of critical, exactly integrable, $S=1$ spin chain 
(Takhtajan-Babujian point) with the central charge $c=3/2$. The dimerization pattern emerging after crossing the critical line is the following: the chains become dimerized in a staggered way to each other with a nonzero relative dimerization. This is consistent with the fact that for the antiferromagnetic interchain interaction effective $S=1$ spins exhibiting nonzero string order are formed across the ladder diagonals rather than along the rungs. On either side of
this line the system is gapped, described in terms of free massive
Majorana fermions with the symmetry $SU(2)\otimes Z^2$.  The gap
(which is the mass of the Majorana triplets) opens linearly as one deviates from the criticality. Owing to the $SU(2)$ symmetry of the model no other
perturbations than mass terms of Majoranas are allowed. In the weak-coupling limit the existence of gapless excitations other than on this line is
thus excluded.

We argue that this quantum critical line (as well as the phases it separates) determined in the weak-coupling limit is smoothly connected to the one discovered in the strong rung coupling limit. The existence of the lowest triplet excitation gap which vanishes linearly as we approach the critical line from the rung dimerized side of the phase diagram (which is seen in both the weak coupling and the strong coupling limit) is taken as evidence that this part as well as the critical line are smoothly connected in weak and strong coupling limit. 

\begin{figure*}
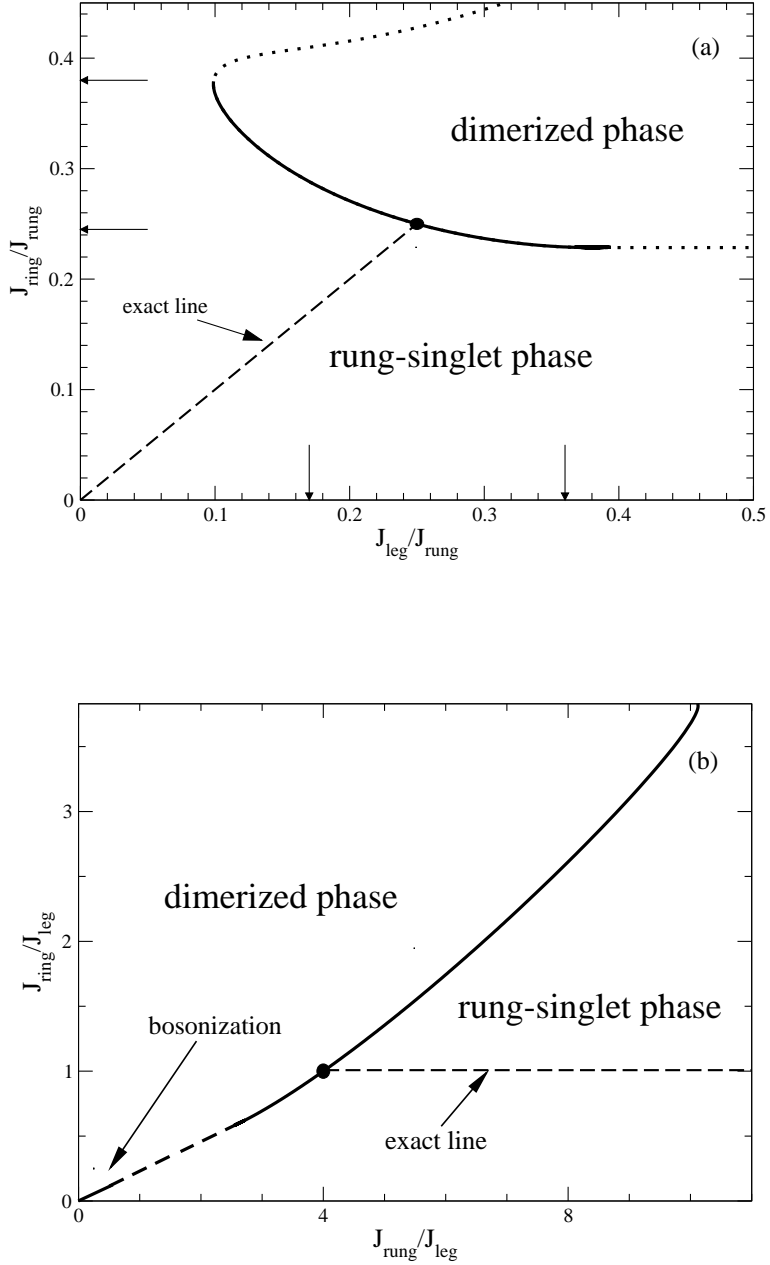

\includegraphics[scale=0.45]{RingPhasendiagramm_en.eps}\\
\vspace*{18mm}
\includegraphics[scale=0.45]{PhaseRingVRung.eps}
\caption{\label{fig:phase} Phase diagrams: (a) result from the 14$^{\rm th}$ order of
perturbation theory in $\ar=J_{\rm ring}/J_{\rm rung}$ and $\al=J_{\rm leg}/J_{\rm rung}$. The arrows mark the regions where the perturbation theory is considered to be valid, while the dotteded and dashed parts of the phase boundary indicate the regions where the cluster method loses convergence; (b) the same in $(J_{\rm ring}$,$J_{\rm rung})$ parameter space showing the smooth connection between bosonization results at small and numerical results at large $J_{\rm ring}$ and  $J_{\rm rung}$.}
\end{figure*}

A closer inspection of our results shows that the smooth connection of the transition line in the two regimes is verified quantitatively: the numerical data shown in Fig.~\ref{fig:phase}(a) seem to indicate asymptotic behavior of the phase boundary in the limit of large $J_{\rm leg}$,
\[
\displaystyle\lim_{J_{\rm leg}\to\infty}J_{\rm ring}^{\rm c}/J_{\rm rung}^{\rm c}\approx 0.22\,.
\]
This number is consistent with the data for the critical line obtained from Lanczos exact diagonalization\cite{MMueller99} for $J_{\rm leg}/J_{\rm rung}\approx 1$ and for finite system size $(N=24)$.
In an alternative presentation of $J_{\rm ring}$ vs.  $J_{\rm rung}$ (see Fig.~\ref{fig:phase}(b)) this asymptotic behavior translates into a straight line joining smoothly the result following \eqref{eq:mass}, $J_{\rm ring}^{\rm c} \sim J_{\rm rung}^{\rm c}$. Thus the limiting value $J_{\rm ring}^{\rm c}/J_{\rm rung}^{\rm c} \approx 0.22$ as obtained in strong coupling appears at the same time a numerical determination of $\alpha^{-1}$ in \eqref{eq:mass}.

Based on that the critical line is specified in terms of three massless Majoranas: the $SU(2)_2$ Wess-Zumino model is the universality class describing this line throughout the phase diagram (consistent with numerical results \cite{Hijii01}). The gapless behavior arises due to fine tuning of rung and ring exchange parameters which leads to mass cancellation of the Majoranas. Upsetting of this fine tuning on either side of the critical line will give rise to nonzero Majorana masses.  

\section{Conclusion}

In summary, we have studied the isotropic $S=1/2$ antiferromagnetic
spin ladder with four-spin ring exchange, both in the limit of weakly
coupled chains and in the strongly coupled (dimer) limit.

In the dimer limit, the use of the linked cluster expansion technique
has allowed us to calculate the dispersion of the elementary
excitations to a high order in the perturbation parameters $J_{\rm
leg}$ and $J_{\rm ring}$, and to establish the shape of the transition
line where the gap vanishes linearly.  In the opposite limit of weakly
coupled chains, using the continuum-limit bosonized form of the
Hamiltonian, we have come to the conclusion that the gap behaves
linearly at both sides of the transition. We identify the phase which
emerges above the critical value of $J_{\rm ring}$ as a spontaneously
dimerized ``non-Haldane'' phase \cite{Nersesyan97} whose elementary
excitations are pairs of massive kinks. From our two complementary
perturbation approaches a consistent description of the unique
transition line in the full phase diagram has emerged. However our
methods cannot access the regions where $J_{\rm rung} \approx J_{\rm
leg}$ and $J_{\rm ring} \gtrsim J_{\rm rung},\,J_{\rm leg}$. Therefore
it is not excluded that some other phases emerge within those regions
as indicated by DMRG calculations\cite{Hikihara02,Troyer02}.

\begin{acknowledgments}

We are grateful to A.~K.~Kolezhuk for numerous
fruitful discussions. One of us (T.V.) is supported by the
Graduiertenkolleg ``Quantum field theory Methods in Elementary
Particles, Gravity and Statistical Physics'' at Hannover
University.

\end{acknowledgments}



\begin{figure*}
\includegraphics[scale=0.45]{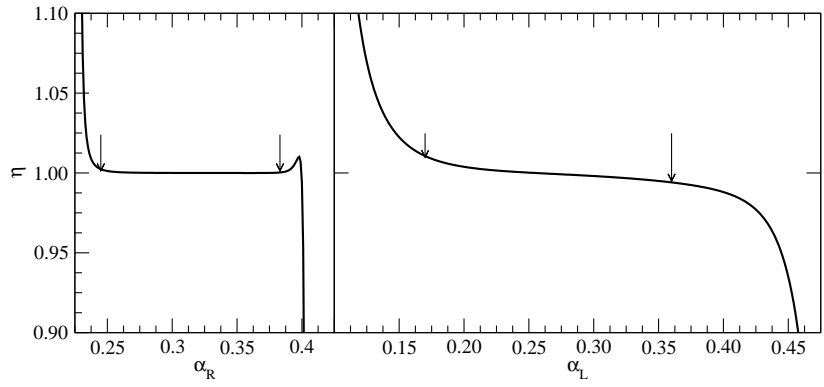}
\caption{\label{fig:exp} The critical exponent 
$\eta$ as a function of $\al$ resp. $\ar$ using a $\log^2 [6,6]$ Pad\'e approximant.} 
\end{figure*}

\end{document}